\documentclass[useAMS,usenatbib]{aa}

\usepackage[varg]{txfonts}
\usepackage{graphicx,color}

\begin{document}

\title{
Detailed chemical compositions of the wide binary HD 80606/80607: revised stellar properties and constraints on planet formation\thanks{The data presented herein were obtained at the W.M.\ Keck Observatory, which is operated as a scientific partnership among the California Institute of Technology, the University of California and the National Aeronautics and Space Administration. The Observatory was made possible by the generous financial support of the W.M.\ Keck Foundation.}}

\author{
F. Liu\inst{1}\thanks{E-mail: fan.liu@astro.lu.se}
\and D. Yong\inst{2}
\and M. Asplund\inst{2}
\and S. Feltzing\inst{1}
\and A.J. Mustill\inst{1}
\and J. Mel\'endez\inst{3}
\and I. Ram\'irez\inst{4}
\and J. Lin\inst{2}}

\institute{
Lund Observatory, Department of Astronomy and Theoretical physics, Lund University, Box 43, SE-22100 Lund, Sweden
\and Research School of Astronomy and Astrophysics, Australian National University, Canberra, ACT 2611, Australia
\and Departamento de Astronomia do IAG/USP, Universidade de Sao Paulo, Rua do Matao 1226, Sao Paulo 05508-900, SP, Brasil
\and Tacoma Community College, Washington, USA}

\date{Received 25 Jan 2018 / Accepted 25 Feb 2018}

\abstract{Differences in the elemental abundances of planet-hosting stars in binary systems can give important clues and constraints about planet formation and evolution. In this study we performed a high-precision, differential elemental abundance analysis of a wide binary system, HD 80606/80607, based on high-resolution spectra with high signal-to-noise ratio obtained with Keck/HIRES. HD 80606 is known to host a giant planet with the mass of four Jupiters, but no planet has been detected around HD 80607 so far. We determined stellar parameters as well as abundances for 23 elements for these two stars with extremely high precision. Our main results are that (i) we confirmed that the two components share very similar chemical compositions, but HD 80606 is marginally more metal-rich than HD 80607, with an average difference of $+$0.013 $\pm$ 0.002 dex ($\sigma$ = 0.009 dex); and (ii) there is no obvious trend between abundance differences and condensation temperature. Assuming that this binary formed from material with the same chemical composition, it is difficult to understand how giant planet formation could produce the present-day photospheric abundances of the elements we measure. We cannot exclude the possibility that HD 80606 might have accreted about 2.5 to 5 $M_{\rm Earth}$ material onto its surface, possibly from a planet destabilised by the known highly eccentric giant.}


\titlerunning{Detailed chemical compositions of HD 80606/80607}
\authorrunning{Liu et al.}
\maketitle

\section{Introduction}

It is well known that stars with higher metallicity have a higher probability of hosting giant planets (e.g. \citealp{gon97,fv05,us07}). However, the effect of metallicity on terrestrial planet formation is poorly understood. \citet{mel09} found that the Sun, which hosts several rocky planets, is depleted in refractory elements when compared to the majority of solar twins, and the differences in elemental abundances correlate with the condensation temperature ($T_{\rm cond}$). The authors proposed that the correlation is likely due to terrestrial planet formation assuming that forming rocky planets in the solar system locked up solid material during the accretion stage so that the process leads to a deficiency of refractories relative to volatiles in the solar photosphere. Studies by \citet{ram09} and \citet{ram10} favour this hypothesis. \citet{liu16a} also found similar signatures of terrestrial planet formation when comparing Kepler-10, which hosts at least one rocky planet, to its stellar twins. Other studies argued that these subtle differences in elemental abundances could be due to Galactic chemical evolution \citep{adi14}, different stellar ages \citep{nis15}, or the dust-cleansing hypothesis: some dust in the pre-solar nebula was radiatively cleansed by luminous hot stars in the solar neighbourhood before the formation of the Sun and its planets. The latter hypothesis is supported by studies of M67, which is an open cluster showing similar elemental abundances to that in the solar photosphere \citep{one11,one14,liu16b}. Alternatively, post-formation accretion of inner planets can also alter the stellar photospheric abundances. If the host stars are polluted after their birth by inner refractory-rich planetary material, the convective envelope of the stars may be enhanced in high-$T_{\rm cond}$ elements (e.g. \citealp{pin01}). Such a process can produce the $T_{\rm cond}$-dependent trend showing enrichment of elemental abundances for refractory elements in the planet host star. 

Binary stars are assumed to form from the same molecular cloud and share identical chemical composition. This is supported both by observations (e.g. \citealp{vog12,kin12}) and by numerical simulations (e.g. \citealp{kra11,rm12}). A differential analysis of the chemical compositions of binary stars can thus reveal the possible signatures related to planet formation or accretion regardless of Galactic chemical evolution. Planet-hosting binary systems are thus the most ideal targets for testing these possible hypotheses related to planet formation.

Recently, high-resolution spectroscopic studies have been applied to several binary systems hosting at least one planet or debris disc. These studies show varied results. \citet{des04,des06} reported that a few pairs of binaries have differences in [Fe/H] between $\approx$ 0.03 to 0.1 dex, although other elements in addition to iron were not studied in detail. Differences in elemental abundances between two components that correlate with $T_{\rm cond}$ were found in 16 Cygni A/B \citep{tuc14} and XO-2 N/S \citep{ram15}. A similar trend was detected in another system (HD 20807/20766) where one of the components hosts a debris disc \citep{saf16}. These three studies favour the scenario proposed by \citet{mel09}. Meanwhile, \citet{oh17} and \citet{saf17} reported two cases where the planet host star in the binary system shows an enrichment in elemental abundances at the level of $\sim$ 0.2 dex for HD 240430/240429 and $\sim$ 0.1 dex for HAT-P-4 system, respectively. These two studies instead favour the scenario of later accretion of inner planets. Moderate abundance differences without clear $T_{\rm cond}$ trend were reported in two cases, WASP-94 A/B \citep{tes16a} and HD 133131 A/B \citep{tes16b}, but no conclusive interpretation was offered. Studies of other binary systems, such as the HAT-P-1 system \citep{liu14}, HD 20781/20782 \citep{mac14}, and HD 80606/80607 (\citet[][hereafter S15]{saf15}; \citet[][hereafter M16)]{mac16} showed no clear abundance differences between the two components, which indicates that harbouring close-in giant planets does not necessarily alter the stellar photospheric abundances. We note, however, that the pair HD 20781/20782 is not ideal for precise abundance characterization due to a large difference in temperature of about 500 K. To summarize, based on the current observational results, it remains unclear whether and how the formation of different types of planets would affect the chemical compositions of their host stars.

HD 80606/80607 is a co-moving wide binary system with an angular separation of $\sim$ 20.6" and a projected separation of $\sim$ 1200 AU \citep{rag06}. Both stars are solar-type with the same spectral type (G5V) and similar apparent V magnitude of 9.06 and 9.17, respectively \citep{kha01}. HD 80606 hosts a giant planet with a mass of $\sim$ 4 $M_{\rm Jup}$ on a very eccentric (e $\sim$ 0.93) orbit at a distance $\sim$ 0.5 AU from its host star \citep{nae01,pon09}. The very eccentric orbit of this planet is probably due to the influence of HD 80607 \citep{wm03}. Meanwhile, such an eccentric warm Jupiter could have interacted with the inner planets, if any existed, and scattered or pushed them to hit the host star \citep{mus15,mus17}, and it might have altered the stellar photospheric abundances. This system therefore provides us with an ideal test case to understand the effect of a close-in Jupiter with high eccentricity on the metallicity of its host star. Spectroscopic studies with uncertainties in differential abundances $<$ 0.05 dex were addressed by S15 and M16. In this paper, we report an independent, detailed spectroscopic analysis of this system with improved precision ($<$ 0.01 dex) in differential abundances, as well as an extended suite of elements examined, using spectra with higher resolution (R = $\lambda/\delta\lambda$) and higher signal-to-noise ratio (S/N), in order to carefully revise the stellar properties and elemental abundances of HD 80606/80607 and to give constraints on close-in giant planet formation and evolution. Furthermore, we performed a dynamical study to explore one way in which HD 80606 could have been polluted by a planet.

\section{Observations and data reduction}

\begin{figure}
\centering
\resizebox{\hsize}{!}{\includegraphics[width=\columnwidth]{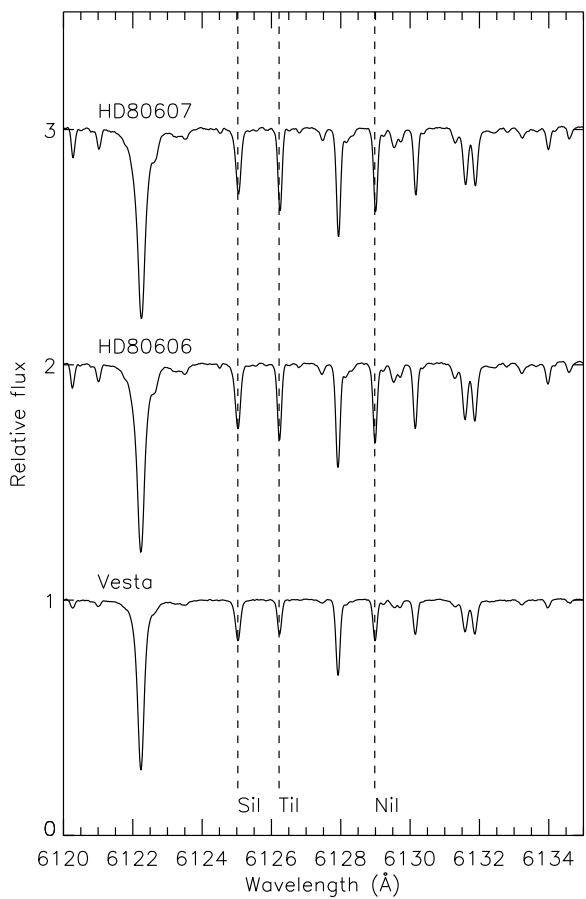}}
\caption{Portion of the normalized spectra for the programme stars HD 80606, HD 80607, and the solar spectrum reflected off the asteroid Vesta. A few atomic lines (Si\,{\sc i}, Ti\,{\sc i}, and Ni\,{\sc i}) used in our analysis in this region are indicated by the dashed lines.}
\label{fig1}
\end{figure}

We obtained high-resolution (R = 86,000) spectra of HD 80606/80607 using the 0.4" slit and 'kv408' filter of the High Resolution Echelle Spectrometer (HIRES, \citealp{vog94}) on the 10 m Keck I telescope on 1$^{\rm{}}$ November 2015. The spectral wavelength coverage is almost complete from 420 to 860 nm. We observed HD 80606 and HD 80607 on the same night with the same spectrograph configuration. The total exposure times were 1800s and 1900s for HD 80606 and HD 80607, respectively. We made use of the Keck-MAKEE pipeline for standard echelle spectra reduction, including bias subtraction, flat-fielding, scattered-light subtraction, spectra extraction, wavelength calibration, and barycentric velocity corrections. We made radial velocity corrections for our spectra and then normalized and co-added the spectra with IRAF\footnote{IRAF is distributed by the National Optical Astronomy Observatory, which is operated by Association of Universities for Research in Astronomy, Inc., under cooperative agreement with National Science Foundation.}. The combined spectrum for each component achieved very high S/N ($\sim$ 400 per reduced pixel). A solar spectrum with S/N $\sim$ 500 per pixel was obtained by observing the asteroid Vesta on the same night with the same instrumental configuration, and the spectrum was reduced using the same method as for the stars. A portion of the reduced spectra for each of the programme stars and for Vesta is shown in Figure \ref{fig1}. We note that the quality of our spectra, which is proportional to R $\times$ (S/N) at a given wavelength \citep{nor01}, is about 60\% better than the previous data used by both S15 and M16 (R $\sim$ 70,000 and S/N $\sim$ 300).

The line-list includes spectral lines of 22 elements (C, O, Na, Mg, Al, Si, S, Ca, Sc, Ti, V, Cr, Mn, Fe, Co, Ni, Cu, Zn, Sr, Y, Ba, and Ce). Most of the data in the list are adopted from \citet{sco15a,sco15b} and \citet{gre15}, and the data were complemented with additional unblended lines from \citet{mel12} and \citet{ben14}. Equivalent widths (EWs) were measured manually via the \textit{splot} task in IRAF, using Gaussian profile fitting. We note that each spectral line was measured consecutively for all the programme stars by setting a consistent continuum, resulting in precise measurements in a differential sense. Strong lines with EW $\geq$ 120 m\AA\ were excluded from the analysis to limit the effects of saturation with the exception of a few Mg\,{\sc i}, Mn\,{\sc i,} and Ba\,{\sc i} lines. The atomic line data as well as the EW measurements we adopted for our analysis are listed in Table A1. We emphasize that in a strictly line-by-line differential abundance analysis of nearly identical stars, the adopted atomic data (e.g. $gf$-values) have very little effect on the results (see e.g. \citealp{liu14}).

\begin{figure}
\centering
\resizebox{\hsize}{!}{\includegraphics[width=\columnwidth]{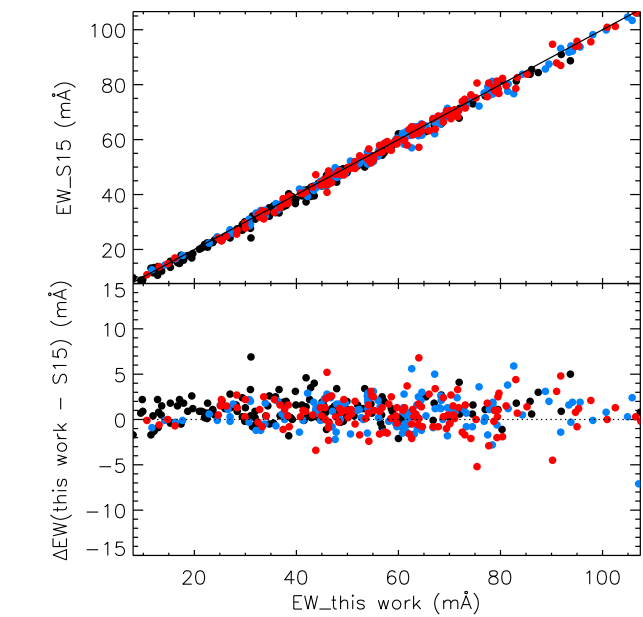}}
\caption{Comparison of EW measurements. Top panel: Comparison of EW measurements between this work and the work by S15. The black, blue and red filled circles represent the EW measurements of the lines in common for the Sun, HD 80606, and HD 80607, respectively. The black solid line represents the one-to-one relation. Bottom panel: Similar to the top panel, but showing the difference in EW measurements.}
\label{fig2}
\end{figure}

\begin{figure}
\centering
\resizebox{\hsize}{!}{\includegraphics[width=\columnwidth]{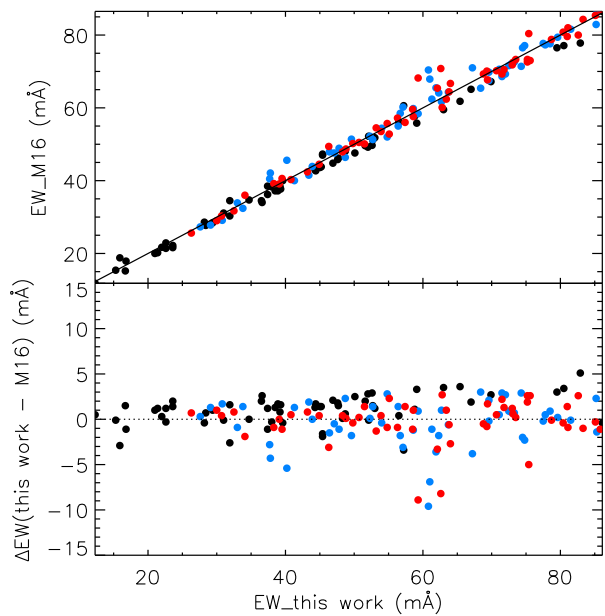}}
\caption{Comparison of EW measurements. Same as Figure \ref{fig2}, but for this work versus M16.}
\label{fig3}
\end{figure}

For lines in common between our study and S15 and M16, we compared the EW measurements. There are 142 lines in common with the work by S15. The mean differences are $<$EW$_{\rm this\,work} $ $-$ EW$_{\rm S15}>$ = 1.07 $\pm$ 1.34 m\AA\ for the Sun, 0.78 $\pm$ 1.90 m\AA\ for HD 80606, and 0.66 $\pm$ 1.91 m\AA\ for HD 80607. When compared to the work by M16, whith which we have 59 lines in common, we found that the differences are $<$EW$_{\rm this\,work} $ $-$ EW$_{\rm M16}>$ = 1.01 $\pm$ 1.63 m\AA\ for the Sun, $-$0.32 $\pm$ 2.64 m\AA\ for HD 80606 and $-$0.25 $\pm$ 2.37 m\AA\ for HD 80607. The scatter is slightly larger when comparing our results to those by M16. Figures \ref{fig2} and \ref{fig3} show the comparison of EW measurements between this work and the work by S15 and M16, respectively. When comparing our measurements to those from M16 for the Sun, we saw a trend (0.04 $\pm$ 0.01) with a dispersion of 1.5 m\AA\ (as shown in the bottom panel, Figure \ref{fig3}). We note that in the work by M16, their solar spectrum was not taken during the same run as when they observed HD 80606 and HD 80607. This might lead to a subtle systematic offset in the EW measurements, as discussed in \citet{bed14}, who found systematic differences in the analysis of solar spectra taken from different instruments or different epochs. We note that there are no systematic offsets between our measurements and the measurements for HD 80606 and HD 80607 by both S15 and M16, and that the smaller scatter between S15 and our work suggests a higher precision in both works.

\section{Analysis and results}

\subsection{Stellar atmospheric parameters}

\subsubsection{Differential stellar parameters}

We performed a 1D local thermodynamic equilibrium (LTE) elemental abundance analysis using the 2014 version of MOOG \citep{sne73,sob11} with the ODFNEW grid of Kurucz model atmospheres \citep{ck03}. Stellar atmospheric parameters (i.e. effective temperature $T_{\rm eff}$, surface gravity $\log g$, microturbulent velocity $\xi_{\rm t}$, and metallicity [Fe/H]) were obtained by forcing excitation and ionization balance of Fe\,{\sc i} and Fe\,{\sc ii} lines on a strictly line-by-line basis relative to the Sun. The adopted parameters for the Sun are $T_{\rm eff} = 5772$\,K, $\log g$ = 4.44 [cm\,s$^{-2}$], $\xi_{\rm t}$ = 1.00 km\,s$^{-1}$, and [Fe/H] = 0.00. The stellar parameters of HD 80606 and HD 80607 were established separately using the automatic grid-searching technique described by \citet{liu14}. In general, the best combination of $T_{\rm eff}$, $\log g$, $\xi_{\rm t}$, and [Fe/H], minimizing the slopes in [Fe\,{\sc i}/H] versus lower excitation potential (LEP) and reduced EW ($\log$\,(EW/$\lambda$) as well as the difference between [Fe\,{\sc i}/H] and [Fe\,{\sc ii}/H] was obtained from a successively refined grid of stellar atmospheric models. The final solution was determined when the step-size of the grid decreased to $\Delta T_{\rm eff}$ = 1 K, $\Delta \log g$ = 0.01 [cm\,s$^{-2}$] and $\Delta$$\xi_{\rm t}$ = 0.01 km\,s$^{-1}$. We also forced the derived average [Fe/H] to be consistent with the adopted value from the stellar atmospheric model. Iron lines whose abundances departed from the average in the final result by $> 3\,\sigma$ were clipped.

\begin{figure}
\centering
\resizebox{\hsize}{!}{\includegraphics[width=\columnwidth]{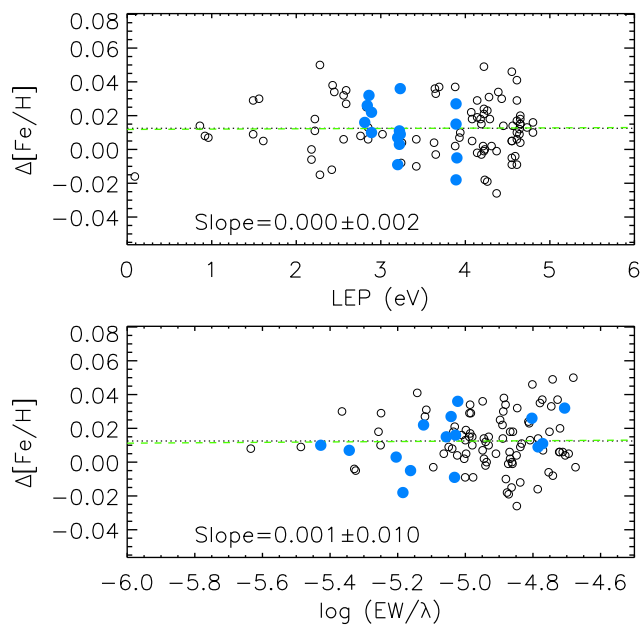}}
\caption{Example of a differential excitation and ionization balance. Top panel: $\Delta$[Fe/H] of HD 80606 derived on a line-by-line basis with respect to HD 80607 as a function of LEP; open circles and blue filled circles represent Fe\,{\sc i} and Fe\,{\sc ii} lines, respectively. The black dotted line shows the location of the mean difference in [Fe/H], and the green dashed line represents the linear least-squares fit to the data. The coefficient for the slope in the best fit and its error is given in the panel. Bottom panel: Same as in the top panel, but as a function of reduced EW.}
\label{fig4}
\end{figure}

We note that HD 80606/80607 are more metal-rich than the Sun by $\sim$ 0.3 dex. Therefore the Sun is not an ideal reference star in terms of a differential analysis. We also derived the differential stellar parameters of the planet host component of HD 80606 relative to HD 80607 using the same method. The adopted parameters for the reference star HD 80607 are $T_{\rm eff} = 5506$\,K, $\log g$ = 4.38 [cm\,s$^{-2}$], $\xi_{\rm t}$ = 0.95 km\,s$^{-1}$, and [Fe/H] = 0.30. Figure \ref{fig4} shows an example of determining the differential stellar atmospheric parameters of HD 80606 relative to HD 80607. The adopted stellar parameters satisfy the excitation and ionization balance in a differential sense.

The final adopted stellar atmospheric parameters of HD 80606/80607 with respect to the Sun, as well as the differential stellar parameters of HD 80606 relative to HD 80607, are listed in Table \ref{t:para}. The stellar parameters for HD80607 are essentially identical when using either the Sun or HD80606 as the reference. The adopted uncertainties in the stellar parameters were calculated using the method described by \citet{eps10} and \citet{ben14}, which accounts for the co-variances between changes in the stellar parameters and the differential iron abundances. Extremely high precision was achieved as a result of the high-quality spectra and the strictly line-by-line differential method we used, which greatly reduces the systematic errors from atomic line data and short-comings in the 1D LTE modelling of the stellar atmospheres and spectral line formation (see e.g. \citealp{asp05,asp09}). The uncertainties in the stellar parameters for HD 80606/80607 with respect to the Sun are $\approx$ 13 - 15 K for $T_{\rm eff}$, 0.03 [cm\,s$^{-2}$] for $\log g$, and 0.02 dex for [Fe/H]. The uncertainties in the differential stellar parameters for HD 80606 relative to HD 80607 are even smaller: $\approx$ 5 K for $T_{\rm eff}$, 0.014 [cm\,s$^{-2}$] for $\log g$, and 0.007 dex for [Fe/H].

\begin{table*}
\caption{Stellar atmospheric parameters.}
\centering
\label{t:para}
\begin{tabular}{@{}lcccc@{}}
\hline\hline
Object & $T_{\rm eff}$ & $\log g$ & [Fe/H] & $\xi_{\rm t}$ \\
 & (K) & (cm\,s$^{-2}$) & (dex) & (km/s) \\
\hline
Vesta$^{a}$ & 5772 & 4.44 & 0.0 & 1.0 \\
HD 80606 & 5584 $\pm$ 13 & 4.40 $\pm$ 0.03 & 0.316 $\pm$ 0.018 & 1.01 $\pm$ 0.04 \\
HD 80607 & 5506 $\pm$ 15 & 4.38 $\pm$ 0.03 & 0.303 $\pm$ 0.018 & 0.95 $\pm$ 0.05 \\
\hline\hline
HD 80607$^{a}$ & 5506 & 4.38 & 0.30 & 0.95 \\
HD 80606 & $+$76 $\pm$ 5 & $+$0.02 $\pm$ 0.01 & $+$0.013 $\pm$ 0.007 & $+$0.06 $\pm$ 0.02 \\
\hline
\end{tabular}
\\
$^a$ Adopted stellar parameters for the reference star.
\end{table*}

\subsubsection{Comparison to the previous studies}

The stellar atmospheric parameters derived in this work, S15, and M16 are listed in Table \ref{t:comp} for comparison. We find that our derived stellar parameters agree with those reported in S15 and M16 within the uncertainties. However, we note that our results are more similar to those of S15, but differ slightly from M16. Our stellar parameter uncertainties are smaller by a factor of 2 - 2.5 when compared to the work by S15 and M16 because our spectra have a higher S/N. Although our more precise results are mainly due to spectra of better quality that result in better measurements, our critical selection of lines, that is, avoiding the use of strong lines and possibly blended lines, might also help to yield more precise stellar parameters. In addition, we used the stellar parameters from M16 to derive the iron abundances using our own line-list. We found clear trends ($>$ 5\,$\sigma$) between the Fe\,{\sc i} abundances as a function of reduced EWs as well as LEP. By adjusting $\xi_{\rm t}$, we cannot achieve the excitation balance based on the lines adopted in our analysis. Considering these facts, we confirm that HD 80606 is marginally more metal-rich than HD 80607 in [Fe/H] by $\approx$ 0.013 dex, which was also reported by S15.

\begin{table*} 
\caption{Comparison of stellar atmospheric parameters$^a$ derived in this work, S15, and M16.}
\centering
\label{t:comp}
\begin{tabular}{@{}lcccc@{}}
\hline\hline
Object & $T_{\rm eff}$ & $\log g$ & [Fe/H] & $\xi_{\rm t}$ \\
 & (K) & (cm\,s$^{-2}$) & (dex) & (km/s) \\
\hline
HD 80606, this work & 5584 $\pm$ 13 & 4.40 $\pm$ 0.03 & 0.316 $\pm$ 0.018 & 1.01 $\pm$ 0.04 \\
HD 80606, S15 & 5573 $\pm$ 43 & 4.32 $\pm$ 0.14 & 0.330 $\pm$ 0.005 & 0.89 $\pm$ 0.09 \\
HD 80606, M16 & 5613 $\pm$ 44 & 4.43 $\pm$ 0.08 & 0.35 $\pm$ 0.02 & 1.36 $\pm$ 0.07 \\
\hline\hline
HD 80607, this work & 5506 $\pm$ 15 & 4.38 $\pm$ 0.03 & 0.303 $\pm$ 0.018 & 0.95 $\pm$ 0.05 \\
HD 80607, S15 & 5506 $\pm$ 21 & 4.31 $\pm$ 0.11 & 0.316 $\pm$ 0.006 & 0.86 $\pm$ 0.17 \\
HD 80607, M16 & 5561 $\pm$ 43 & 4.47 $\pm$ 0.06 & 0.35 $\pm$ 0.02 & 1.26 $\pm$ 0.07 \\
\hline
\end{tabular}
\\
$^a$ Relative to the Sun.
\end{table*}

\subsection{Abundance results}

\subsubsection{Differential elemental abundances}

We performed the 1D LTE analysis with the MOOG 2014 version to determine elemental abundances of 22 elements in addition to Fe (C, O, Na, Mg, Al, Si, S, Ca, Sc, Ti, V, Cr, Mn, Co, Ni, Cu, Zn, Sr, Y, Ba, La, and Ce) by fitting a curve of growth from EW measurements of lines in the spectra of HD 80606/80607 and the Sun. Hyperfine structure splitting (HFS) corrections were considered for Sc, V, Cr, and Cu, with the HFS data taken from \citet{kur95}. Considering that the Mn lines used in our study might be affected by blending as discussed in \citet{fel07}, we also applied the spectral synthesis approach to test the Mn abundances using the data from \citet{pro00} and a similar method as discussed in \citet{bb15} for two lines in common (Mn\,{\sc i} 6013\AA\, and Mn\,{\sc i} 6016\AA). We found that the results derived from spectral synthesis agree well with those derived from EW measurements\footnote{The relatively similar results obtained using different methods probably arise because we studied stars that are very similar. These results do not mean that HFS can be ignored.}. Although the absolute values obtained for each star vary by $\sim$ 0.05 dex using different methods, the differential abundances for HD 80606/80607 relative to the Sun as well as for HD 80606 relative to HD 80607 do not change essentially. Then we derived the chemical abundances of HD 80606/80607 relative to the Sun adopting a strictly line-by-line approach. For differential elemental abundances of HD 80606 relative to HD 80607, we adopted the differential stellar parameters rather than using the stellar parameters with respect to the Sun for consistency and precision.

We adopted 3D NLTE corrections for the oxygen abundance determination from the 777nm triplet based on \citet{ama16}. The differential 3D NLTE abundance correction (HD 80606 $-$ HD 80607) for oxygen is $\approx -$0.012 dex, which should not be neglected. We also estimated the differential 3D NLTE abundance corrections for Al and Si for HD 80606 relative to HD 80607.\\
\textit{Aluminium.} 3D NLTE corrections for two Al lines (669.6nm and 669.8nm) were adopted from \citet{nl17}. The differential 3D NLTE abundance correction for Al is about $+$0.001 to $+$0.002 dex.\\
\textit{Silicon.} 3D NLTE corrections for our Si lines were calculated based on the grid provided by Amarsi (priv. comm.), where the radiative transfer code and model atom were presented in \citet{aa17}. The average differential 3D NLTE abundance correction for Si is $\approx -$0.001 dex.\\
We obtained the differential 1D NLTE abundance corrections for Na, Mg, Ca, Ti, Cr, Mn, Fe, Co, Cu, and Ba for HD 80606 relative to HD 80607 as described below.\\
\textit{Sodium.} 1D NLTE corrections were calculated based on \citet{lin11} for four Na lines (475.1nm, 514.8nm, 615.4nm, and 616.0nm) using the INSPECT database\footnote{Version 1.0: www.inspect-stars.com.}. The differential 1D NLTE abundance correction for Na is $\approx +$0.001 dex.\\
\textit{Magnesium.} The differential 1D NLTE abundance correction for one line (571.1nm) is $+$0.002 dex, derived using the INSPECT database, based on \citet{ob16}. For another two Mg lines used in this work (631.8nm and 631.9nm), the differential 1D NLTE abundance corrections are both zero, calculated using the GUI web-tool from Maria Bergemann's group\footnote{Latest version: http://nlte.mpia.de/gui-siuAC\_sec.php.} based on \citet{ber15}. The average differential 1D NLTE abundance correction for Mg is $\approx +$0.001 dex.\\
\textit{Calcium.} 1D NLTE corrections for four Ca lines (616.6nm, 645.5nm, 647.1nm, and 649.9nm) were estimated using the grid from \citet{mas17}. The differential 1D NLTE abundance correction for Ca is $\approx -$0.001 dex.\\
\textit{Titanium.} 1D NLTE corrections for our Ti lines were derived using the GUI web-tool based on \citet{ber11}. The average differential 1D NLTE abundance correction for Ti is $\approx +$0.002 dex.\\
\textit{Chromium.} 1D NLTE corrections for several Cr lines used in this work were adopted from \citet{bc10}, as well as the GUI web-tool. The differential 1D NLTE abundance correction for Cr is about $+$0.001 to $+$0.002 dex.\\
\textit{Manganese.} We calculated the 1D NLTE corrections for four Mn lines (500.4nm, 603.1nm, 601.6nm, and 602.1nm) using the GUI web-tool based on \citet{bg08}. The differential 1D NLTE abundance correction for Mn is $\approx -$0.001 dex.\\
\textit{Iron.} 1D NLTE corrections for our Fe lines were derived using the INSPECT database based on \citet{lin12}. The average differential 1D NLTE abundance correction for Fe is zero.\\
\textit{Cobalt.} 1D NLTE corrections for three Co lines (521.2nm, 645.4nm, and 741.7nm) were calculated using the GUI web-tool based on \citet{ber10}. The differential 1D NLTE abundance correction for Co is $\approx +$0.001 dex.\\
\textit{Copper.} We estimated the 1D NLTE correction for one line (521.8nm) based on the grid from \citet{yan15} and found that the differential 1D NLTE abundance correction for Cu is about $+$ 0.003 to $+$ 0.004 dex.\\ 
\textit{Barium.} We derived the 1D NLTE corrections for three Ba lines (585.3nm, 614.1nm, and 649.6nm), adopted from \citet{kor15}. The average differential 1D NLTE abundance correction for Ba is $\approx -$0.002 dex.\\
We note that except for oxygen, NLTE corrections for our differential abundances between HD 80606 and HD 80607 are almost negligible (see Table \ref{t:abun}, Col. 4), which should not affect our main results essentially. However, the NLTE corrections can be more prominent when comparing HD 80606/80607 to the Sun because of a $T_{\rm eff}$ difference of more than 200 K and a [Fe/H] difference of $\sim$ 0.3 dex. Therefore we only focus on the differential abundances for HD 80606 relative to HD 80607 in the following discussions, although the elemental abundances of HD 80606/80607 relative to the Sun are still listed in Table \ref{t:abun} for reference.

We found that the average abundance difference of all species in this wide binary system (HD 80606 $-$ HD 80607) is $+$0.013 $\pm$ 0.002 dex with a standard deviation of $\sigma$ = 0.009 dex. Our results agree well with S15, who reported that HD 80606 is slightly more metal-rich than HD 80607 ($+$0.010 $\pm$ 0.019 dex). However, their results were not as statistically significant because of the larger errors, while our results confirm their finding for almost all elements thanks to our improved precision. We note that our results from the differential abundance analysis do not agree with the work by M16, who reported that HD 80606 is slightly more metal-poor than HD 80607 ($-$0.018 $\pm$ 0.004 dex) based on 13 refractory elements ($T_{\rm cond}$ $>$ 900 K).

Errors in differential abundances in Table \ref{t:abun} correspond to the standard error of the line-to-line scatter added in quadrature combined with the error introduced by the uncertainties in the stellar atmospheric parameters. The process of our error analysis is similar to the method employed by \citet{eps10} and \citet{ben14}, as mentioned before. The differential elemental abundances for most species between HD 80606 and HD 80607 have uncertainties smaller than 0.01 dex (see Table \ref{t:abun}). The average error in differential abundances is only $\approx$ 0.007 dex when comparing HD 80606 to HD 80607. We improved the precision of the abundance analysis significantly when compared to the previous studies by S15 and M16.

\begin{table*}
\caption{Differential elemental abundances.}
\centering
\label{t:abun}
\begin{tabular}{@{}ccccc@{}}
\hline\hline
Species & HD 80606$^a$ & HD 80607$^a$ & $\Delta$(80606 - 80607)$^b$ & $\Delta$(NLTE) \\
 & (dex) & (dex) & (dex) & (dex) \\
\hline
 C\,{\sc i}  & 0.261 $\pm$ 0.015 & 0.256 $\pm$ 0.014 &  0.006 $\pm$ 0.006 & ... \\
 O\,{\sc i}  & 0.252 $\pm$ 0.027 & 0.264 $\pm$ 0.027 & 0.003 $\pm$ 0.007 & $-$0.012$^c$ \\
Na\,{\sc i}  & 0.403 $\pm$ 0.016 & 0.389 $\pm$ 0.014 &  0.012 $\pm$ 0.005 & $+$0.001$^d$ \\
Mg\,{\sc i}  & 0.316 $\pm$ 0.008 & 0.303 $\pm$ 0.013 &  0.013 $\pm$ 0.006 & $+$0.001$^d$ \\
Al\,{\sc i}  & 0.330 $\pm$ 0.027 & 0.326 $\pm$ 0.028 &  0.003 $\pm$ 0.003 & $+$0.001 - $+$0.002$^c$ \\
Si\,{\sc i}  & 0.344 $\pm$ 0.013 & 0.340 $\pm$ 0.014 &  0.005 $\pm$ 0.003 & $-$0.001$^c$ \\
 S\,{\sc i}  & 0.345 $\pm$ 0.015 & 0.338 $\pm$ 0.014 &  0.008 $\pm$ 0.004 & ... \\
Ca\,{\sc i}  & 0.278 $\pm$ 0.019 & 0.272 $\pm$ 0.019 &  0.005 $\pm$ 0.005 & $-$0.001$^d$ \\
Sc\,{\sc ii} & 0.369 $\pm$ 0.022 & 0.333 $\pm$ 0.024 &  0.035 $\pm$ 0.008 & ... \\
Ti\,{\sc i}  & 0.356 $\pm$ 0.017 & 0.345 $\pm$ 0.019 &  0.009 $\pm$ 0.007 & $+$0.002$^d$ \\
Ti\,{\sc ii} & 0.331 $\pm$ 0.024 & 0.319 $\pm$ 0.023 &  0.012 $\pm$ 0.009 & ... \\
 V\,{\sc i}  & 0.392 $\pm$ 0.027 & 0.382 $\pm$ 0.033 &  0.007 $\pm$ 0.010 & ... \\
Cr\,{\sc i}  & 0.346 $\pm$ 0.017 & 0.336 $\pm$ 0.016 &  0.009 $\pm$ 0.006 & $+$0.001 - $+$0.002$^d$ \\
Mn\,{\sc i}  & 0.421 $\pm$ 0.023 & 0.407 $\pm$ 0.023 &  0.013 $\pm$ 0.005 & $-$0.001$^d$ \\
Fe\,{\sc i}  & 0.316 $\pm$ 0.012 & 0.302 $\pm$ 0.013 &  0.013 $\pm$ 0.005 & 0.0$^d$ \\
Fe\,{\sc ii} & 0.316 $\pm$ 0.021 & 0.305 $\pm$ 0.022 &  0.012 $\pm$ 0.007 & 0.0$^d$ \\
Co\,{\sc i}  & 0.377 $\pm$ 0.023 & 0.364 $\pm$ 0.021 &  0.012 $\pm$ 0.008 & $+$0.001$^d$ \\
Ni\,{\sc i}  & 0.398 $\pm$ 0.012 & 0.386 $\pm$ 0.013 &  0.012 $\pm$ 0.005 & ... \\
Cu\,{\sc i}  & 0.418 $\pm$ 0.013 & 0.390 $\pm$ 0.013 &  0.026 $\pm$ 0.004 & $+$0.003 - $+$0.004$^d$\\
Zn\,{\sc i}  & 0.365 $\pm$ 0.014 & 0.345 $\pm$ 0.022 &  0.021 $\pm$ 0.012 & ... \\
Sr\,{\sc i}  & 0.314 $\pm$ 0.035 & 0.300 $\pm$ 0.036 &  0.011 $\pm$ 0.027 & ... \\
 Y\,{\sc ii} & 0.293 $\pm$ 0.033 & 0.269 $\pm$ 0.033 &  0.025 $\pm$ 0.011 & ... \\
Ba\,{\sc ii} & 0.213 $\pm$ 0.032 & 0.196 $\pm$ 0.033 &  0.017 $\pm$ 0.009 & $-$0.002$^d$ \\
Ce\,{\sc ii} & 0.269 $\pm$ 0.032 & 0.243 $\pm$ 0.034 &  0.026 $\pm$ 0.007 & ... \\
\hline
\end{tabular}
\\
$^a$ Differential 1D LTE elemental abundances [X/H] relative to the Sun.\\
$^b$ Differential 1D LTE elemental abundances $\Delta$[X/H] of HD 80606 relative to HD 80607.\\
$^c$ Differential 3D NLTE abundance corrections for HD 80606 relative to HD 80607.\\
$^d$ Differential 1D NLTE abundance corrections for HD 80606 relative to HD 80607.\\
\end{table*}

\begin{figure}
\centering
\resizebox{\hsize}{!}{\includegraphics[width=\columnwidth]{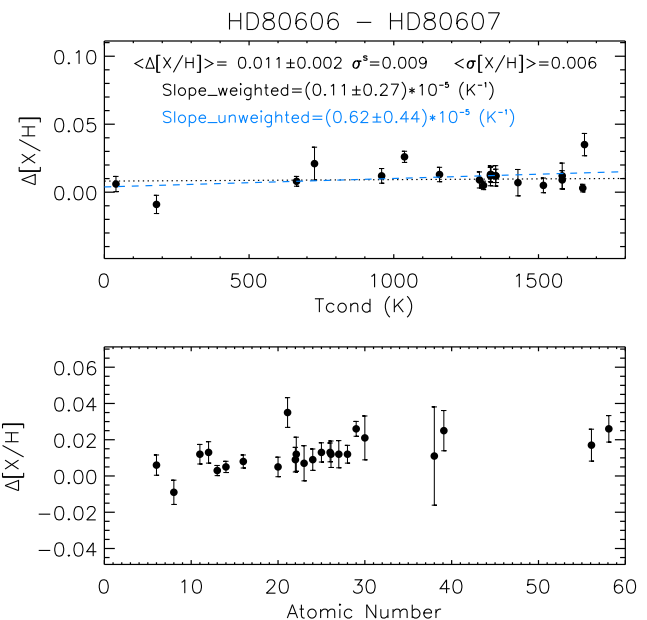}}
\caption{Differential elemental abundances of HD 80606 relative to HD 80607. Top panel: Differential elemental abundances of HD 80606 relative to HD 80607 as a function of $T_{\rm cond}$ (Z $\leq$ 30). The black and blue dotted lines show the weighted and unweighted linear least-squares fits to the data, respectively. Bottom panel: Differential elemental abundances of HD 80606 relative to HD 80607 as a function of atomic number for all the elements.}
\label{fig5}
\end{figure}

Our main results are plotted in Figure \ref{fig5}, showing the differential elemental abundances of HD 80606 relative to HD 80607 as a function of $T_{\rm cond}$ (upper panel) and atomic number (lower panel). The values of 50\% $T_{\rm cond}$ were taken from \citet{lod03}. The correlation between elemental abundances and $T_{\rm cond}$ might help us to probe the possible signatures imprinted by planet formation \citep{mel09,saf17}. We only include elements with atomic number Z $\leq$ 30 when analysing the trend between differential abundances ($\Delta$[X/H]) and $T_{\rm cond}$ because heavy elements with Z $>$ 30 might be affected by other factors such as $s$-process and $r$-process production (see e.g. \citealp{mel12,liu16b}). We emphasize that differential abundances with extremely high precision were derived with an average uncertainty of $\approx$ 0.007 dex. We recall that the average abundance difference is $+$0.013 $\pm$ 0.002 dex ($\sigma$ = 0.009 dex), indicating that HD 80606, the star with a giant planet, is marginally more metal-rich than HD 80607 with about 6 -- 7\,$\sigma$ significance for most elements. The slope of the $\Delta$[X/H] -- $T_{\rm cond}$ trend is (0.11 $\pm$ 0.27) $\times 10^{-5}$\,K$^{-1}$ for a weighted linear least-squares fitting and (0.62 $\pm$ 0.44) $\times 10^{-5}$\,K$^{-1}$ for an unweighted linear least-squares fitting to the data. This shows that no $T_{\rm cond}$ trend of the differential abundances between HD 80606 and HD 80607 was detected in our analysis.

\subsubsection{Comparison to the previous studies}

\begin{figure}
\centering
\resizebox{\hsize}{!}{\includegraphics[width=\columnwidth]{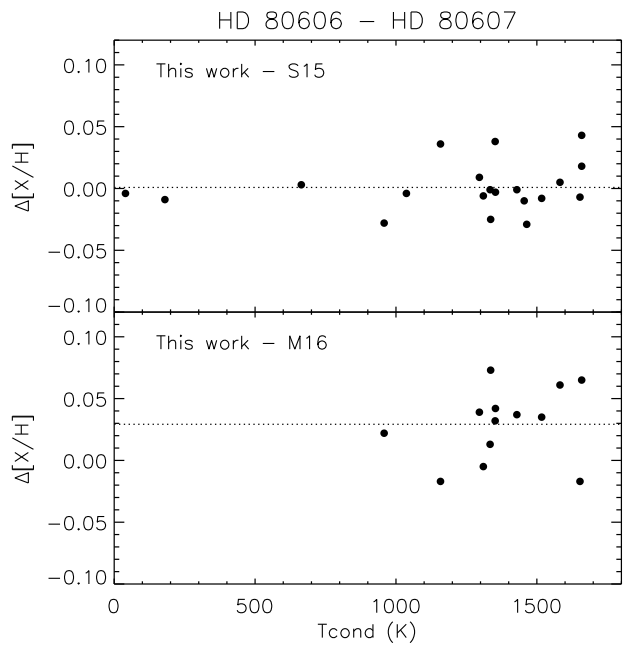}}
\caption{Comparison of the abundances from this work to the previous studies. Top panel: Abundance differences in $\Delta$[X/H] (HD 80606 $-$ HD 80607) for elements in common in our work and the work by S15. The dashed lines indicate the location for the mean values of abundance differences. Bottom panel: Similar as in the top panel, but showing the comparison between this work and the work by M16.}
\label{fig6}
\end{figure}

We compared our derived abundances to the previous studies by S15 and M16 for elements in common. For the elements in common in this work and S15, the average abundance differences are $-$0.014 $\pm$ 0.104 dex for HD 80606 relative to the Sun, $-$0.015 $\pm$ 0.105 dex for HD 80607 relative to the Sun, and 0.001 $\pm$ 0.020 dex for HD 80606 relative to HD 80607. The scatters are smaller when using the differential abundances between these two components, and no clear systematic offsets in differential abundances (HD 80606 $-$ HD 80607) were detected. When compared to the work by M16, with which we have 13 elements in common, we found the average abundance differences to be $-$0.024 $\pm$ 0.031 dex for HD 80606 relative to the Sun, $-$0.058 $\pm$ 0.045 dex for HD 80607 relative to the Sun, and 0.029 $\pm$ 0.029 dex for HD 80606 relative to HD 80607. For differential abundances (HD 80606 $-$ HD 80607), the results of M16 are systematically higher by $\sim$ 0.03 dex. Fig \ref{fig6} shows the abundance differences for HD 80606 relative to HD 80607 in each element in common in our work and the studies by S15 and M16, respectively. We note that the average uncertainty in the abundance differences between HD 80606 and HD 80607 is only $\approx$ 0.007 dex in this work, which is about 3 -- 4 times smaller than the differences in the works of S15 and M16 (0.020 dex and 0.027 dex, respectively), emphasizing the importance of higher quality spectra.

\subsection{Stellar ages}

We derived the stellar ages and masses for HD 80606/80607 using the \textit{q$^{2}$} Python package developed by I. Ram\'irez\footnote{https://github.com/astroChasqui/q2.}, which employs the theoretical Yonsei-Yale isochrones \citep{dem04}. We estimated the stellar ages using different combinations of stellar parameters and list two cases below.
\begin{itemize}
\item Using purely spectroscopic $T_{\rm eff}$ and $\log g$. The age of HD 80607, the reference star in our differential analysis, is 7.5 $\pm$ 1.0 Gyr. Then we adopted the differential stellar parameters for HD 80606 for its age estimation, which essentially decreased the relative uncertainty. The age of HD 80606 is 6.0 $\pm$ 0.5 Gyr. We note that the ages of the two components are not consistent, with a difference of $\sim$ 1.5 Gyr in this case.
\end{itemize}
\begin{itemize}
\item Using spectroscopic $T_{\rm eff}$ but the parallaxes from Gaia Data Release 1 \citet{gai16} to determine the ages. The parallaxes are 15.22 $\pm$ 0.28 mas for HD 80607 and 15.33 $\pm$ 0.26 mas for HD 80606. The ages were estimated to be 5.8 $\pm$ 1.0 Gyr for HD 80607 and 4.7 $\pm$ 0.6 Gyr for HD 80606. In this case, the ages of the two components are marginally inconsistent, with a difference of $\sim$ 1.1 Gyr. 
\end{itemize}
In both cases, we found that the stellar masses are $\approx$ 1.03 $M_{\rm Sun}$ for HD 80606 and $\approx$ 1.00 $M_{\rm Sun}$ for HD 80607.

We now proceed to a discussion of the measured ages. Assuming that HD 80606 and HD 80607 formed together, their ages cannot be different. This differs from what we find in this work: $\sim$ 1 -- 1.5 Gyr difference in age with about 2 -- 3\,$\sigma$ significance. One possibility is that the age difference might arise from erroneous spectroscopic $\log g$. We find that slight changes in spectroscopic $\log g$ ($\sim$ 0.01 dex) do not alter the elemental abundances (i.e. metallicity difference) essentially but will affect the age determination, which is sensitive to the location in the $T_{\rm eff}$ -- $\log g$ diagram. 

Another interesting possibility for explaining the difference in age is that one of the programme stars, HD 80606, has had its elemental abundances changed because it has engulfed a planet. According to \citet{dc03} and \citet{cs05}, the impact of pollution from planet engulfment on stellar evolution models can change the evolution of the stars in the Hertzsprung-Russell diagram and hence change the derived age. If we assume that the convection zone of HD 80606 has been enriched through accretion of an inner planet, it should be warmer than its original temperature based on the polluted stellar evolution models. Therefore the detected temperature difference could be due to the event of planet engulfment, which leads to the age difference. For a further discussion of the plausibility of this scenario, see Section 4.3.

\subsection{Lithium abundances}

Lithium abundances were determined using the spectral synthesis method with the 2014 version of MOOG, as described in \citet{car16}. In short, we first estimated the stellar broadening using clean lines around 600 nm, and with the instrumental and stellar broadening fixed, we fit the lithium region using the line list of \citet{mel12}. NLTE corrections for lithium were computed using the grid from \citet{lin09}. Finally, the errors were estimated based on uncertainties in the continuum setting, the deviations of the observed profile relative to the synthetic spectrum, and the uncertainties introduced by the errors in the stellar parameters.

For HD 80606, we calculated that the macroturbulent velocity (v$_{\rm macro}$) is 2.62 km/s and v$_{\rm sini}$ is 1.80 $\pm$ 0.33 km/s. We then obtained A(Li)$_{\rm NLTE}$ = 0.87$\pm$ 0.05 dex and the LTE value to be 0.77 $\pm$ 0.05 dex. For HD 80607, we determined that v$_{\rm macro}$ = 2.42 km/s, v$_{\rm sini}$ = 1.69 $\pm$ 0.28 km/s, A(Li)$_{\rm NLTE}$ = 0.80 $\pm$ 0.05 dex and the LTE value to be 0.70 $\pm$ 0.05 dex. While the lithium abundance is slightly low for solar age and mass, it is reasonable considering that the metallicity of HD 80606/80607 is much higher than solar values, therefore the convective zone should be deeper than in the Sun \citep{cas09}. We note that the abundance of Li is slightly higher in HD 80606 than in HD 80607, which might be due to the slightly higher mass of HD 80606 compared to HD 80607, although we cannot exclude the possibility that HD 80606 is enriched by accretion of an inner planet due to the effect of the current eccentric giant planet (see Section 4.3). 

\section{Discussion}

\subsection{Relevant hypotheses of the abundance differences}

The chemical compositions of HD 80606/80607 may provide some constraints on the planet formation in the wide binary system, although we cannot exclude the existence of a planet around HD 80607. There are several ways in which the presence of planets can change the composition of elements in a stellar atmosphere. Here we mainly discuss two possibilities. The first (which we call the depletion hypothesis) explains the deficit of refractory elements in a planet-hosting star as compared to an otherwise similar star without a planet as the results of the planet formation process; the star lacks these elements as they have been tied up in the planets. This hypothesis is based on the assumption that forming a terrestrial planet removed solid material from the proto-planetary disc that was accreted later by the host star, leading to a deficiency in stellar abundances for refractory elements. This hypothesis was first put forward in \citet{mel09} and has been further discussed in \citet{ram09}, \citet{ram10}, and \citet{liu16a}. In principle, a planet-hosting star should be depleted in refractories, showing a negative trend between elemental abundances and $T_{\rm cond}$ when compared to stellar twins without planets.

Another possibility to explain the differences in elemental abundances in stars with and without planets was put forward by \citet{pin01}. In this scenario, the abundances of the planet-hosting star might be enhanced in refractory elements due to the post-formation accretion of inner planets/material. This scenario is based on the assumption that the presence of a close-in giant planet might scatter or otherwise perturb the inner planets onto the surface of their host star, thus adding a large amount of H-depleted material, enhancing the stellar surface chemical compositions for refractory elements. We refer to this scenario as the enrichment hypothesis. Further discussions and results have been provided by \citet{mac14}, M16, \citet{oh17}, and \citet{saf17}. In general, a planet-hosting component in a binary system could be enriched in refractories, showing a positive trend between elemental abundances and $T_{\rm cond}$ when compared to the other component in this binary system.

\subsection{Abundance results for HD 80606 and HD 80607}

Our results show the overall abundance differences to be 0.013 $\pm$ 0.002 dex ($\sigma$ = 0.009 dex) for almost all elements between HD 80606 and HD 80607, where the planet host star HD 80606 is marginally more metal-rich. However, we found no $T_{\rm cond}$ trend with the differential abundances between the two components. This is different from what has been found in 16 Cygni A/B, where the planet-hosting component is more metal-poor than the component without a known planet. A natural explanation for the results of 16 Cygni A/B is that the giant planet removed metals from the proto-planetary disc that were later accreted by the host star, causing an overall deficiency in metallicity of the host star \citep{ram11,tuc14}. A possible explanation for our results is that HD 80606/80607 might have thicker convection zones than 16 Cygni A/B, while the chemical signature of planet formation might be washed out during the accretion and depletion phase, but it is also possible that because of the small abundance differences we cannot detect a trend with $T_{\rm cond}$.

Considering that the planet around HD 80606 is very eccentric (e $\sim$ 0.93) and orbits at $\sim$ 0.5 AU \citep{nae01,pon09}, the fact that HD 80606 is slightly more metal-rich than HD 80607 by $\sim$ 0.01 dex probably favours the enrichment scenario in which the inner planets were ingested onto the planet host star HD 80606 through the effects of this known eccentric planet. This scenario merits further investigation.

\subsection{Exploration of the enrichment scenario}

We tested the possibility of the proposed enrichment scenario through dynamical simulations. The known planet HD 80606 b may be undergoing tidal circularisation from an initially wider configuration, after its eccentricity was forced to high values by Kozai cycles from the binary companion HD 80607 \citep{ft07}. This high-eccentricity migration of a proto-hot Jupiter is highly destabilising for any planets that exist close to the star, as they are strongly perturbed when the proto-hot Jupiter acquires its high eccentricity and small perihelion. The dominant outcome of instabilities in such a system is collision of the inner planet(s) with the star \citep{mus15,mus17}, which could account for the observed enrichment.

\begin{figure}
\centering
\resizebox{\hsize}{!}{\includegraphics[width=\columnwidth]{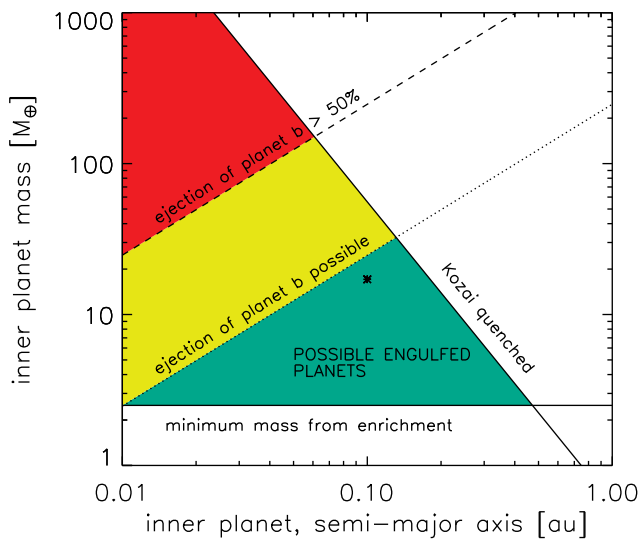}}
\caption{Constraints in semi-major axis and mass for the possible engulfed planet. The stellar abundances set a minimum mass of 2.5 $M_{\rm Earth}$. For a given semi-major axis, a hard limit on the possible planet mass is set by the requirement that it not suppress Kozai cycles on the outer planet b. A soft limit is set by the increasing probability of ejecting planet b as the engulfed planet's orbital energy becomes more negative.}
\label{fig7}
\end{figure}

\begin{figure}
\centering
\resizebox{\hsize}{!}{\includegraphics[width=\columnwidth]{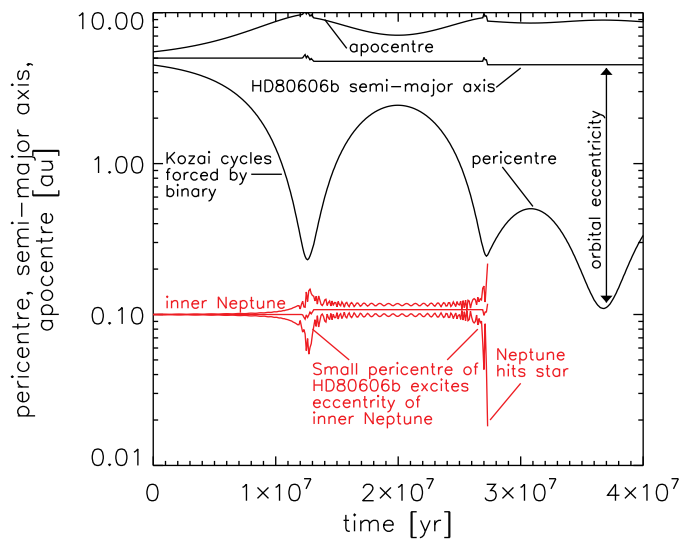}}
\caption{Dynamical simulation of the possible influence of HD 80606 b to the inner Neptune-mass planet. The black and red lines show the pericentre, semi-major axis, and apocentre of HD 80606 b and the engulfed planet, respectively. The binary companion forces Kozai cycles on planet b, which perturbs the inner planet when its pericentre is small and forces the inner planet into the star during the second such cycle.}
\label{fig8}
\end{figure}

What can we say about the properties of the possible engulfed planet? In addition to the minimum possible mass of the engulfed planet obtained from the abundance measurements, there are dynamical constraints setting maximum masses for the engulfed planet. If the planet had been too massive and/or too close to planet b, it would have induced a precession on the orbit of planet b sufficient to quench Kozai cycles and prevent the excitation of the observed high eccentricity \citep{mus17}. Furthermore, a planet with a too high orbital binding energy has a chance of ejecting planet b from the system, instead of being forced into the star \citep{mus15}. These constraints demarcate a region in semi-major axis and mass for the engulfed planet illustrated in Figure \ref{fig7}. Here the minimum mass required for enrichment and the maximum mass to avoid Kozai quenching are shown as hard limits, while the maximum mass to avoid ejection of planet b is shown as a soft limit with gradations in shading marking where ejection becomes possible and likely. We took the initial parameters from \citet{ft07} for the state of the binary and planet b. Under these assumptions, the engulfed planet must have been interior to $\sim$ 0.5 AU, with no other large planets between it and the initial orbit of planet b at 5 AU. As an example, we numerically integrated systems with a Neptune-mass planet placed at 0.1 AU (marked with a star in Figure \ref{fig7}). Of 100 systems, 93 saw their inner planet collide with the star between 10 and 100 Myr into the integration. One example is shown in Figure \ref{fig8}: HD 80606 b experiences large eccentricity oscillations due to the Kozai cycles imposed by the binary, which in turn excite the eccentricity of the interior Neptune, and in the second cycle at $\sim$ 27 Myr, force it into the star.

\begin{figure}
\centering
\resizebox{\hsize}{!}{\includegraphics[width=\columnwidth]{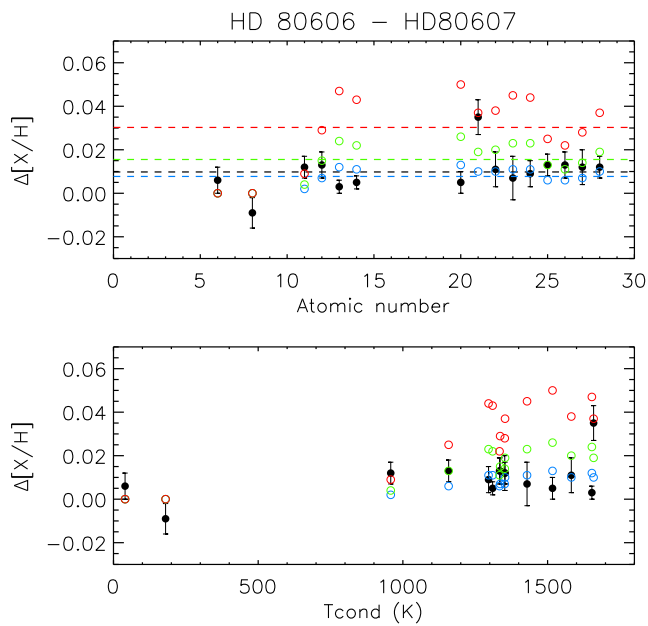}}
\caption{Differential abundances for (HD 80606 $-$ HD 80607) as a function of atomic number (top panel) and $T_{\rm cond}$ (bottom panel). The black points represent our results. The blue, green, and red points are the model-predicted values corresponding to the effect of 2.5 $M_{\rm Earth}$, 5 $M_{\rm Earth}$ and 10 $M_{\rm Earth}$, respectively (M16). The black, blue, green, and red dashed lines indicate the locations for the mean values of the corresponding data points.}
\label{fig9}
\end{figure}

Assuming that this scenario is true, we compared the observed abundance differences between HD 80606 and HD 80607 to the predicted values that correspond to the ingested material based on the method described by \citet{cha10}, \citet{mac14}, and M16. We found that for refractory elements, the abundance differences between HD 80606 and HD 80607 correspond to $\sim$ 2.5 to 5 $M_{\rm Earth}$ material. Figure \ref{fig9} shows how much material was accreted onto HD 80606 when compared to HD 80607 if we assume that the overall abundance differences between these two components are due to the enrichment of inner planets. However, it is hard to explain why the volatile elements are also enriched by a similar level, unless we were to assume that a Neptune-like planet with lots of volatiles and ices was accreted. No $\Delta$[X/H] -- $T_{\rm cond}$ trend was detected between these two components, which further complicates and challenges the enrichment scenario.

\subsection{Summary and prospect}

To summarize, we found that HD 80606 is marginally more metal-rich than HD 80607 by 0.013 $\pm$ 0.002 dex ($\sigma$ = 0.009 dex). We assume that subtle differences in abundances ($\sim$ 0.01 dex) are highly unlikely to be due to inhomogeneities in the proto-stellar cloud; stars in a binary system are presumably born from a chemically homogeneous gas cloud. In addition, we note that almost all the chemical elements are enhanced in HD 80606 relative to HD 80607 instead of random variations implied by an inhomogenous medium. We performed a T-test to estimate the probability that the abundance differences (HD 80606 $-$ HD 80607) arise by chance. The control sample for our test was a sample of zeros, representing the null abundance difference between the two components. We obtained a T-value of 6.79 and a probability $p = 1.85 \times 10^{-8}$ to obtain such a high T-value, which indicates that it is extremely unlikely that the abundance difference of these stars is 0. We also repeated the t-test 100,000 times by randomly varying the abundances based on their uncertainties and obtained the average T-value to be 5.06 and $p = 0.0005$. Thus, even considering the errors, it is still unlikely that the observed abundance difference could arise solely by chance.

The observed abundance differences might favour the enrichment scenario, as discussed above. However, we note that the small 0.01 dex difference could be the combined effect of both planet formation and planet evolution (migration and planet engulfment) around both components. Considering the important fact that no $T_{\rm cond}$ trend of the differential elemental abundances between HD 80606 and HD 80607 was detected, it is difficult to explain how giant planet formation could produce the present-day elemental abundances. It is also possible that the presence of a close-in giant planet does not imprint significant chemical signatures in the photosphere of its host star, as for the HAT-P-1 binary system \citep{liu14}. The observed abundance difference in this work is smaller than in other studies with planet-hosting binaries (e.g. \citealp{tes16a,tes16b}). Our results are contrary to what has been proposed in the case of 16 Cygni A/B \citep{ram11,tuc14} and XO-2 N/S \citep{ram15}. This might be due to different types and planetary orbits that these binary systems host, different timescales for planet formation relative to the accretion of gas, and the size of the convective zone.

A comprehensive and homogeneous study of a large number of planet-hosting binaries with different types of planets (terrestrial planets, hot giants, and cool giants) with high-precision differential analysis is needed in order to explore the possible connections between abundance differences (also $T_{\rm cond}$ trend) in planet-hosting binaries and the properties of planets (e.g. mass, semi-major axis, and eccentricity). Such a study will provide us with significant constraints on the planet formation process with statistical support. In addition, Gaia DR2 will release a large number of co-moving solar-type binaries that will be ideal targets for studying the possible abundance differences and for testing the proposed depletion or enrichment scenarios.

\section{Conclusions}

We performed a high-precision differential analysis of the wide binary HD 80606/80607. HD 80606 hosts an eccentric giant planet with $\sim$ 4 $M_{\rm Jup}$ mass and orbits at $\sim$ 0.5 AU \citep{nae01,pon09}, while no planet has been detected around HD 80607. We determined the stellar atmospheric parameters of both components relative to the Sun, as well as the differential stellar parameters of HD 80606 relative to HD 80607, with very high precision. Our stellar parameters agree with those reported by S15 and M16, but our uncertainties are a factor of 2 - 2.5 smaller. We also derived stellar ages and masses for HD 80606/80607 using the Yonsei-Yale isochrones with different combinations of stellar parameters.

We determined the elemental abundances of 22 elements for HD 80606/80607 with respect to the Sun, as well as the differential abundances between these two components. We are able to achieve extremely high precision in differential elemental abundances ($\approx$ 0.007 dex). We also derived the lithium abundances for HD 80606/80607, which are both smaller than the solar value. This is most likely due to the higher metallicity ($\sim$ 0.3 dex) of this binary system. The abundance of Li in HD 80606 is slightly higher than HD 80607, probably because of the slightly different masses.

We confirm that HD 80606 is marginally more metal-rich than HD 80607 by a factor of $\sim$ 0.01 dex. The average abundance difference (HD 80606 $-$ HD 80607) is $+$0.013 $\pm$ 0.002 dex with a standard deviation of $\sigma$ = 0.009 dex. The average difference in abundances is significant at $\sim$ 6 -- 7\,$\sigma$ level. For the correlation between $\Delta$[X/H] and $T_{\rm cond}$, we found no $T_{\rm cond}$ trend of the differential abundances when comparing these two components to each other, but this could be due to the small abundance differences, making it difficult to find any $T_{\rm cond}$ slope. Considering the fact that HD 80606 hosts a close-in giant planet while no known planet has been detected around HD 80607, our results challenge the explanation of the abundance pattern found in 16 Cygni A/B \citep{ram11,tuc14} and XO-2 N/S \citep{ram15}.

We note that assuming the scenario that the photosphere of HD 80606 is enhanced by accretion of inner planets through the interaction of its close-in eccentric giant planet, the observed abundance differences of refractory elements correspond to $\sim$ 2.5 to 5 $M_{\rm Earth}$ of high $T_{\rm cond}$ material. Dynamical simulation shows that the current eccentric hot Jupiter likely forced the inner planet, if any existed, to its host star. However, this scenario cannot fully explain the enhancement in volatile elements in the photosphere of HD 80606. Several factors (e.g. the size of the convection zone, planet formation and evolution, and the composition of the possible engulfed planet) might affect the photospheric abundances we measure, which complicates the scenario for the observed abundance pattern.

Our detailed study of the wide binary HD 80606/80607 showed that a high-precision differential analysis in binary systems with planets can provide important constraints for understanding the effect of planet formation on the chemical compositions of the host stars. Further efforts are needed for a large sample of stellar binaries hosting terrestrial planets and giant planets at different orbits with high-precision differential spectroscopic analyses, in order to systematically study whether and how the formation, evolution, and migration of different types of planets would affect the properties of their host stars.

\section*{Acknowledgments}
F.L. and S.F. acknowledge support by the grant "The New Milky Way" from the Knut and Alice Wallenberg Foundation. F.L. was also supported by the Swedish Research Council (grant 2012-2254). This work has been supported by the Australian Research Council (grants FL110100012, FT140100554 and DP120100991). J.M. acknowledges support by FAPESP (2012/24392-2). A.J.M is supported by the grant "Impact" from the Knut and Alice Wallenberg Foundation.
We thank Luca Casagrande and Christian Sahlholdt for discussions about the determination of photometric parameters and stellar ages. We thank Anish Amarsi for discussions about 3D NLTE corrections of the differential elemental abundances.
The authors thank the ANU Time Allocation Committee for awarding observation time to this project. The authors wish to acknowledge the very significant cultural role and reverence that the summit of Mauna Kea has always had within the indigenous Hawaiian community.

\section*{SUPPLEMENTARY MATERIAL}

The following material is available online for this article:

Table A1. Atomic line data and the EW measurements adopted for our analysis.

\end{document}